\documentclass[12pt,preprint]{aastex}

\begin{document}
\title{{\it GALEX} and Optical Observations of GW Librae During the Long
Decline from Superoutburst\footnote{Based
on observations made with the NASA Galaxy Evolution Explorer and with the
Apache Point Observatory (APO) 3.5m telescope. GALEX is operated
for NASA by the California Institute of Technology under NASA contract 
NAS5-98034. APO is owned and operated by the Astrophysical Research Consortium (ARC).}}
\author{Eric Bullock\altaffilmark{2},
Paula Szkody\altaffilmark{2},
Anjum S. Mukadam\altaffilmark{2},
Bernardo W. Borges\altaffilmark{3},
Luciano Fraga\altaffilmark{4},
Boris T. G\"ansicke\altaffilmark{5},
Thomas E. Harrison\altaffilmark{6},
Arne Henden\altaffilmark{7},
Jon Holtzman\altaffilmark{6},
Steve B. Howell\altaffilmark{8},
Warrick A. Lawson\altaffilmark{9},
Stephen Levine\altaffilmark{10, 11},
Richard M. Plotkin\altaffilmark{2,12},
Mark Seibert\altaffilmark{13},
Matthew Templeton\altaffilmark{7},
Johanna Teske\altaffilmark{14},
Frederick J. Vrba\altaffilmark{10}}
\altaffiltext{2}{Department of Astronomy, University of Washington, Box 351580,
Seattle, WA 98195; ericb98@u.washington.edu, mukadam@astro.washington.edu,
szkody@astro.washington.edu}
\altaffiltext{3}{Faculdade de Ci\^{e}ncias Exatas e Tecnologia, Universidade 
Federal da Grande Dourados, CEP 79804-070, Dourados, Brazil bernardoborges@ufgd.edu.br}
\altaffiltext{4}{Southern Observatory for Astrophysical Research, Casilla 603, La Serena, Chile}
\altaffiltext{5}{Department of Physics, University of Warwick, Coventry CV4 7AL, UK}
\altaffiltext{6}{Department of Astronomy, New Mexico State University, Box
30001, MSC 4500, Las Cruces, NM 88003; holtz@nmsu.edu, tharriso@nmsu.edu}
\altaffiltext{7}{AAVSO, 49 Bay State Road, Cambridge, MA 02138; arne@aavso.org, matthewt@aavso.org}
\altaffiltext{8}{National Optical Astronomy Observatories; 950 North Cherry Avenue, Tucson, AZ 85726; howell@noao.edu}
\altaffiltext{9}{School of Physical, Environmental \& Mathematical Sciences,
University of New South Wales, Australian Defence Force Academy, Canberra ACT 
2600, Australia; w.lawson@adfa.edu.au}
\altaffiltext{10}{US Naval Observatory, Flagstaff Station, Flagstaff, AZ 86001}
\altaffiltext{11}{Lowell Observatory, 1400 West Mars Hill Road, Flagstaff AZ
86001; sel@lowell.edu}
\altaffiltext{12}{Astronomical Institute Anton Pannekoek, University of
Amsterdam, Science Park 904, 1098XH Amsterdam, The Netherlands}
\altaffiltext{13}{Observatories Carnegie Institute of Washington, 813 Santa 
Barbara St. Pasadena, CA 91101, mseibert@obs.carnegiescience.edu}
\altaffiltext{14}{Department of Astronomy, University of Arizona, 933 N. Cherry Ave, Tucson, AZ 85721; jkteske@email.arizona.edu}
\begin{abstract}
The prototype of accreting, pulsating white dwarfs (GW Lib) underwent
a large amplitude dwarf nova outburst in 2007. We used ultraviolet
data from {\it GALEX} and ground-based optical photometry and
spectroscopy to follow GW Lib for three years following this outburst.
Several variations are apparent during this interval. The optical shows a
 superhump modulation in the months following outburst
while a 19 min quasi-periodic modulation lasting for several months is apparent 
in the year after
outburst. A long timescale (about 4 hr) modulation first appears in
the UV a year after outburst and increases in amplitude in the following
years. This variation also appears in the optical 2 years after outburst but
is not in phase with the UV.
The pre-outburst pulsations are not yet visible after 3 years, 
likely indicating the
white dwarf has not returned to its quiescent state.
\end{abstract}

\keywords{cataclysmic variables --- stars: individual (GW Lib) --- 
stars: oscillations}

\section{Introduction}

GW Librae was first thought to be a nova (Duerbeck 1987) because of the large
amplitude of an outburst during its discovery in 1983, but a later spectrum 
at quiescence revealed it to be a
dwarf nova. The lack of outbursts for the next two decades, 
combined with the large outburst
amplitude argued for a classification as a WZ Sge type of dwarf nova
 (Howell, Szkody \&
Cannizzo 1995). Time-resolved spectroscopy (Szkody, Desai \& Hoard 2000; 
Thorstensen et al. 2002) revealed a very short orbital period of 76.78 min,
consistent with this classification.
However, the unique nature of GW Lib became apparent when it was discovered to
have non-radial pulsations (Warner \& van Zyl 1998) and it became the
prototype of the small class of accreting, pulsating white dwarfs which
currently contains 13 members. Exploration of its amplitude spectrum by
van Zyl et al. (2000, 2004) identified three prominent pulsation periods
at 648, 376 and 236 s along with fine structure of closely spaced
frequencies around the 648 and 376 s periods. The amplitudes of the
pulsations changed on monthly timescales, typical of most hydrogen atmosphere
white dwarf pulsators (ZZ Ceti stars). Ultraviolet observations
of GW Lib (Szkody et al. 2002) revealed a hot white dwarf with a temperature
consistent with the theoretical instability strip at 15,000K due
to \ion{He}{2} ionization (Arras et al. 2006). The best fit was achieved with
a high mass (0.8 M$_{\odot}$) white dwarf with 0.1 solar metal abundance
and a dual temperature (63\% of the surface at 13,300K and 37\% at 17,100K).
The UV data showed higher amplitude pulsations than at optical wavelengths 
with the same periods as evident in 
the optical. All stellar pulsators, including variable white dwarfs, reveal
wavelength dependent amplitudes whenever their atmospheres suffer from
limb-darkening effects.

Recent studies of the temperatures of nine of the other accreting pulsators
(Szkody et al. 2010) showed that their instability strip is wider than for
ZZ Ceti stars and some stars appeared to stop pulsating during the
Hubble Space Telescope observations. One possible explanation for at least
one of the systems that did not show pulsations was that accretion heating
from an outburst could have caused the white dwarf 
to move out of
its instability strip. Outbursts are known to cause major heating of
the white dwarfs (e.g. WZ Sge; Godon et al. 2006) which then take several
years to cool to their quiescent values, even though the optical light
reaches quiescence in only a few months.

On 12 April 2007, GW Lib went into outburst (Templeton et al. 2007), reaching
8th magnitude from its quiescent brightness near 17th magnitude. The outburst
was followed spectroscopically (Nogami et al. 2009; Hiroi et al. 
2009; van Spaandonk et al.
2010a) and photometrically (Copperwheat et al. 2009, Schwieterman et al. 2010) 
as well as with
X-ray telescopes (Byckling et al. 2009). The spectroscopy revealed the usual prominent 
optically thick disk near outburst with a slow return toward quiescence over
3 months. However, the outburst was unusual compared to other dwarf novae
and WZ Sge objects
in several ways. The X-ray flux of GW Lib increased by 3 orders of 
magnitude during the outburst and remained one order of magnitude higher than
pre-outburst even at 2 years past outburst. As Bykling et al. (2009) point out,
GW Lib is the only other dwarf nova other than U Gem that shows a larger
X-ray flux during outburst than quiescence. The optical flux of GW Lib 
also remained
about a half-magnitude higher than quiescence 2-3 years past outburst,
whereas most WZ Sge stars return to optical quiescence in a few months
(e.g. AL Com and WZ Sge; Szkody et al. 1998). Finally, there was no evidence of the echo outbursts that many WZ Sge
stars show on the decline from outburst (e.g. EG Cnc; Patterson et al. 2002).

We observed GW Lib for three years following its outburst, using
ultraviolet data from {\it GALEX}, and ground-based optical photometry
and spectroscopy. Our aim was to understand the heating and cooling of
GW Lib as it moved out of its instability strip and then re-entered.
As of the current time, the white dwarf has not yet resumed its
pre-outburst character. Yet, the photometry has revealed some interesting
facets of GW Lib during its decline from superoutburst.

\section{Observations}

\subsection{{\it GALEX} Data}

Ultraviolet data were obtained using the {\it GALEX} satellite in April-June
of 2007, 2008, 2009 and 2010 (Table 1 lists the times). Both the near-ultraviolet
(NUV: 1750-2800\AA) and the far-ultraviolet (FUV; 1350-1750\AA) detectors
(Martin et al. 2005) were used, except for the 2010 data when the FUV detector
was no longer operational. The mean magnitudes from each set
of observations were used to construct a light curve over the three years 
(Figure 1). From the {\it GALEX} online 
documentation\footnote{http:galexgi.gsfc.nasa.gov/tools/index.html}, the
zeropoint of the AB magnitudes is given by m$_{0}$= 18.82=1.40$\times$10$^{-15}$
ergs cm$^{-2}$ s$^{-1}$ \AA$^{-1}$ for FUV and 
 m$_{0}$= 20.08=2.06$\times$10$^{-16}$ ergs cm$^{-2}$ s$^{-1}$ \AA$^{-1}$ for
NUV. 

In order to be able to analyze the data for any short-term periodicity 
related to pulsations,
the time-tag data were binned (into 29s exposures for the 2007 data and 59s for
the 2008-2010 data), calibrated and measured with the IRAF\footnote{IRAF
is distributed by the National Optical Astronomy Observatory, which is
operated by the Association of Universities for Research in Astronomy, Inc.,
under cooperative agreement with the National Science Foundation.} 
{\it qphot} routine. An 8 pixel radius aperture was used for the stars 
and an annulus of 10-13
pixels for sky measurements. Nine reference stars in the field were measured 
in the same way as GW Lib to serve as a check on any variations that could be due to the various
satellite orbits.

\subsection{Optical Photometry}

Optical photometric data were obtained with multiple telescopes between 2007-2010,
beginning four days after the ouburst of GW Lib in April (Templeton et al. 2007). The primary telescope used was the robotic 1.0 meter of New Mexico
State University (NMSU) which utilizes an E2V 2048$\times$2048 CCD. In 2007, the
night typically began with 2-6 observations through $U,B,V,R$ and $I$ filters.
Longer observing runs were then obtained through the $U$ filter. Beginning in
2008, a BG40 
filter\footnote{http://www.us.schott.com/advanced\_optics/english/download/schott\_bandpass\_bg40\_2008\_e.pdf} 
was also used and in 2009, 
only the BG40 filter was
employed. This filter is a Schott glass filter with a broadband response
from 3500-6000\AA\ which allows better S/N than the $U$ filter, 
while still permitting
better detection of pulsations which have higher amplitude in blue
light than in red (Robinson et al. 1995).
 Table 2 lists the ranges of exposures for each night, with the $U$
band images occupying the upper end of the range. Throughout each night, the
exposure times were automatically adjusted to compensate for changing conditions
such as airmass and cloud cover. Differential photometry with respect to
calibrated standard stars in the field was used to obtain magnitudes 
with resulting photometric errors of up to a few hundredths of a magntitude.

On 2007 May 29, a dataset was taken with the 3.5m Wisconsin, Indiana, Yale,
National Optical Astronomy Observatory (WIYN) telescope through a BG39 filter
which has similar transmission to the BG40.
Observations were made with the MiniMo CCD with 2$\times$2 binning and
differential photometry was used for the reductions.

Observations with the 3.5m telescope at Apache Point Observatory (APO) were
made on 2007 June 1 and 2008 March 28. The data were taken through a BG40
filter with the Agile CCD and differential photometry was again used to
obtain the magnitudes.

Data from the 1m telescope at the United States Naval Observatory Flagstaff
Station (NOFS) were also obtained in 2009. A high speed camera was used on
May 15 with a $V$ filter, and the Tektronix 1024$\times$1024 CCD with $B$
filter was used on May 16 and 17. Additionally, the Sonoita Research
Observatory (SRO) 0.5m telescope was used on 2009 May 16. With this telescope,
unfiltered images were taken with an SBIG STL 6303 CCD. Differential
photometry was employed in both reductions.

Southern hemisphere observations were made with the 4.1m Southern Astrophysics 
Research (SOAR) telescope on 2009 May 17. A combination of two E2V CCDs were
used to make measurments through a $V$ filter. Differential photometry using
the IRAF ${\it qphot}$ routine was used to obtain magnitudes. A second southern
dataset was
obtained with the South African Astronomical Observatory 
(SAAO) 1.0 m Elizabeth telescope beginning 2009 May 18. A $B$ filter
was used with the STE4 CCD camera. Differential photometry was then used to
obtain a light curve.

In 2010, two nights of photometry were obtained on March 4 and 5, using the
KPNO 2.1m telescope with a BG39 filter and the STA2 CCD. As with the
other reductions, differential photometry was used.

Approximate calibrated ${\it B}$ magnitudes were obtained for all nights
through a variety of methods. Generally, $B$ and $V$ magnitudes for
comparison stars in the field were obtained from the AAVSO Variable Star
Database and offsets computed to then calibrate the GW Lib lightcurves. 
With the NMSU data, offsets for each night could usually be obtained
from the $B$ filter. For the 2007 nights with no $B$ filter, the offset
was determined by comparison of the mean magnitude  to the corresponding
magnitude from the AAVSO lightcurve. On 2008 Jun 20, data were
taken in both $B$ and BG40 filters, which enabled a determination of 
the offset to transform
the BG40 magnitudes of 2008-2010 to approximate $B$ magnitudes. The
AAVSO standards were also used to obtain a transformation of the clear filter
of SRO to an approximate $B$ magnitude.
To tranform the $V$ filter data of SOAR and NOFS, it was assumed that
the $B-V$ color of GW Lib was zero.

\subsection{Optical Spectroscopy}

An optical spectrum of GW Lib was obtained on 2008 April 17, using the
R-C Spectrograph and grating KPGL1-1 on the CTIO 4m telescope. Two
30 min exposures were medianed together. The slit width was 1 arcsec,
comparable to the seeing (determined from a slice across the spectrum).
The airmass at the time of observation was 1.04 and the blue DA white
dwarf EG274 (Hamuy et al. 1992) was observed at airmass 1.03 immediately
following GW Lib. Due to the low airmass and the closeness in time of
the standard, we estimate the flux uncertainty to be about 10\%. The
resulting calibrated flux at 5500\AA\ translates into a visual magnitude
of 16.3, which is within the range of the AAVSO light curves during 2008.

Table 2 gives a summary
of the dates and times for all the optical data. 

\section{Light Curves}

Figure 1 shows the mean of the ultraviolet and optical blue
 light curves from the observations.
The optical light undergoes a rapid linear decline in the first
month after outburst, followed by a steep drop of 3 magnitudes between May 7-10
(days 227-230 on the plot) and then a slower decline continuing for years
after outburst. 
The steep drop is generally ascribed to large
changes in the disk that bring it close to its pre-outburst state (Byckling
et al. 2009).  The {\it GALEX} coverage began after
this steep drop had occurred. The ultraviolet brightness declines by 1.5 magnitudes from
2007 June to 2008 June, while the optical changes by less than a magnitude
during this time (the low point at day 560 is likely due to using a redder
standard for calibration due to the smaller field of view of the detector
during that observation as well as a different filter). 
The difference between the {\it GALEX} FUV
and NUV magnitudes decreases from 0.4 mag in 2007
to 0.08 mag in 2009. The optical blue magnitude
decreases slightly in 2010 while the ultraviolet NUV magnitude remains the same.
These declines are consistent with the cooling of a
 hot source that peaks in the FUV, while the NUV and optical comprise
the tail of the flux distribution. The similarity of the NUV fluxes
from 2008-2010 indicates the continued presence of a hot source.

Figure 2 shows the individual {\it GALEX} light curves in NUV and FUV
filters for the various satellite orbits on a given date, 
while Figure 3 shows the optical
light curves for the nights with data coverage longer than an hour.

The individual {\it GALEX} light curves are flat throughout the 2008 May and 
June observations.
The optical light curves are also relatively flat from 2007 April
16-22 (Figure 3a), which corresponds to the timescale of the linear decline in 
Figure 1.
The flat light curves are consistent with the domination of the system light by 
the
optically thick uniform accretion disk. On May 7, at the start of the
steep optical decline (Figure 1), the optical light curves
go through major changes (Figure 3b). On May 7, the decline starts to
appear during the 4 hrs of observation. Three nights later, at the
end of the rapid decline on May 10, the light curve shows large
fluctuations of several tenths of a magnitude, with a double-hump appearance
repeating at a timescale near 80 min. On May
14, a slow sinusoidal type modulation is apparent near 80 min that
is also visible on May 29. This is the same interval in which Kato et al. (2008)
find a stable period of 77.98 min which they call a ``late superhump" period.
The erractic fluctuations in our May 10 data may be the disk transitioning from
its early to late superhump state. 
 Kato et al. (2008) note the strange stability and longer
than usual period increase of this superhump compared to normal WZ Sge stars.
Copperwheat et al. (2009) note a similar long period in their
2007 July data but with a lower amplitude modulation
 than we see in May. These longer
period superhumps have been postulated to be due to some amount of gas being
outside the 3:1 resonance radius, which causes a tidal instability leading to 
an eccentric disk (Whitehurst 1988). The lack of modulation of the NUV
and FUV data during this time implies that the inner, hotter disk areas
are not affected. As the superhump amplitude
decreases with time, the amount of mass in this outer area of
the disk may be shrinking or there may be less irradiation of the disk as
time progresses, although
the low inclination of GW Lib ($\sim$ 11$\deg$; van Spaandonk et al. 2010a)
should present a view of the disk that does
not change much during the orbit. It is also intriguing that GW Lib did not
show the series of rebrightening events (echo outbursts) 
following its steep optical decline
that are often observed in WZ Sge
systems (Patterson et al. 2002).

A year later, when GW Lib was again observable, the {\it GALEX} data begin
to show a long (hrs) timescale variation that becomes even more pronounced
in 2009 and 2010. In 2008 and 2009, the amplitude in FUV is larger than the NUV,
implying the innermost hot areas of the disk are involved in the cause
of this variation.
On 2008 March 29 (Figure 3b), a long term modulation that is not well-defined
(as it is about the length of the 3 hr light curve) appears in the optical
data, and there is an even more puzzling feature: a distinct pulse-like periodicity. 
The DFT (Figure 4) indicates a period
of 19.08$\pm$0.03 min.
Copperwheat et al. (2009)
also report a consistent signal from March 30-April 29 at 19.2$\pm$0.2 min
 with an
amplitude in $g$ of 2.2\%, which declines in amplitude throughout their 
April data. In addition, Schwieterman et al. (2010) show a similar (19.7-19.9 min) but not identical period 
in 3 out of 8 nights from 2008 May 30 to June 13. As the frequency is not
stable, we regard this as a quasi-periodic feature that is similar in timescale 
to those that have been reported in several novalike systems
(Garnavich \& Szkody 1992; Taylor et al. 1999; Andronov 1999).

After the disappearance of the 80 min and 19 min phenomena, the
optical light curves in 2008 June and throughout 2009 exhibit a longer modulation
near 4-5 hrs (Figures 3c, 3d) with amplitude of 0.2-0.3 mag peak-to-peak
 which is neither consistent nor coherent.
This timescale is about double the
2.1 hr period 
first discovered by Woudt \& Warner (2002) and then studied further  in the
pre-outburst data of Copperwheat et al. (2009). Figure 3d shows that
while this modulation is very prominent on 2009 May 17, it completely
disappears on May 18 and 19 (as evidenced on data obtained on two
different telescopes). Both {\it GALEX} and optical data exist on 2009 May 16 
(although not simultaneously), showing that a similar modulation is
apparent in both (Figures 2a and 3c and Table 3). However, while the periods
are similar on the same date (the
combined optical data from May 14-17 fit with a period of 191.5$\pm$0.2 min
and the  {\it GALEX} data give periods of 218$\pm$11min (FUV) and 210$\pm$9 min
(NUV)), the phases do not seem to match up. The peak of the {\it GALEX} data
occur near 730-740 UT, which would imply the previous peak occurred at
530-538 UT, while the optical peak is closer to 450 UT. 
While the origin of
this modulation is not understood, the fact that it changes phase on
timescales of days and sometimes disappears entirely, 
points to an origin in the accretion disk. If it is related to an
eccentricity of the disk, the changes mean that the disk is still not
in a steady configuration two years after the superoutburst. 

\section{Pulsation Search}

A period search was conducted on all the optical and ultraviolet data,
primarily to search for the return of the pre-outburst pulsations.
For this, the light curves were converted to a fractional amplitude 
scale by dividing by the mean count rate and then subtracting one.
Then a Discrete Fourier Transform (DFT) was computed for each light
curve up to the Nyquist frequency (half the sampling frequency). 
The best fit periods and amplitudes
were found
by using linear and non-linear least squares fitting. To find the
significance of any period, the 3$\sigma$ white noise limits were
determined by using a shuffling procedure (Kepler, 1993; Mukadam et al. 2010).
This involves shuffling the intensities of a light curve
(with all the periodicities subtracted out), while keeping the times
constant in order to produce a pure white noise lightcurve. The DFT of
this lightcurve is then computed to find the average amplitude between 0
and the Nyquist frequency to yield a 1$\sigma$ measure. Repeating this
process 10 times produces an empirical, reliable 3$\sigma$ limit.

Table 3 lists all the 3$\sigma$ noise levels and any 
significant results obtained for the ultraviolet and optical
data. There is no consistent period that is evident as a non-radial
pulsation up to 3 years past outburst in either the ultraviolet or the optical
data. The amplitudes of the 650s period evident prior to the outburst were
100 mma in the UV and 10-15 mma in the optical. Figures 5 and 6 show the
DFTs of the latest data. 

Copperwheat et al. (2009) also obtained photometry in 2007 June and July
and 2008 Mar-June. In their last dataset (2008 June 21), they note a period
of 290s with an amplitude of 1.25\%. 
There is no consistent
evidence for a pulsation near this period in our longest datasets (Table 3). Thus,
this period is likely associated with flickering noise.

Van Spaandonk et al. (2010b) recently used VLT spectroscopy to determine
a spin period of 97$\pm$12s from an analysis of a \ion{Mg}{2}4481 absorption feature.
In contrast to the obvious presence of the spin period in the accreting 
pulsator V455 And (Araujo-Betancor et al. 2005), there is no indication of this 
spin period in any of our data on GW Lib.

\section{White Dwarf Cooling}

The UV data obtained following the outburst of the dwarf nova
 WZ Sge showed that its white dwarf takes more than
3 yrs to cool to its pre-outburst temperature (Godon et al. 2006).
As with all dwarf novae near the peak of outburst, the UV flux is dominated 
by the accretion
disk and it is only after the flux has dropped down from the plateau
region of the outburst that the white dwarf becomes visible. Nogami et al. (2009),
Hiroi et al. (2009) and van Spaandonk et al. (2010a) 
discuss the spectra of GW Lib from pre-outburst to the slower decline
phase up to 3 months post outburst. Our spectrum (Figure 7) obtained one year after
outburst, is similar to quiescent spectra in showing the absorption from the white
dwarf flanking the emission from the accretion
disk, but there is a steeper blue continuum than at quiescence. As the emission 
lines are relatively
narrow and single-peaked, the inclination of GW Lib is generally taken
to be low ($\sim$11$^{\circ}$) and the mass to be high (near 1 M$_{\odot}$;
 Szkody et al. 2000, Thorstensen et al. 2002, Steeghs et al. 2007, van
Spaandonk et al. 2010a,b).

We modelled the optical spectrum of GW\,Lib using a two-compoment
approach. The emission from the disk is described with an isothermal
and isobaric slab of hydrogen (G\"ansicke et al. 1997).
For the white dwarf, we used a grid of model spectra spanning a
wide range in effective temperatures and surface gravities, computed
using TLUSTY195 and SYNSPEC45 (Hubeny \& Lanz 1995) and adopting metal
abundances of 0.1 the solar values, as determined from our previous
{\it HST}/STIS observations (Szkody et al. 2002), and using the
distance of 104$\pm^{30}_{20}$ as determined by Thorstensen (2003). 
Free parameters in this model are the temperature and column
density of the hydrogen slab, which determine the equivalent widths of
the emission lines as well as the Balmer decrement (see G\"ansicke et
al. 1999 and Rodriguez-Gil et al. 2005
for more details), and the white dwarf effective temperature and
surface gravity. The surface gravity, along with the flux scaling
factor between models and observation and a white dwarf mass-radius
relation (Wood, 1995) give the white dwarf mass.

The disk parameters are adjusted to reproduce the observed H$\beta$
and H$\gamma$ fluxes, and the white dwarf parameters are varied to fit
the broad Balmer absorption lines as well as the slope of the
continuum. As the ionization fraction of hydrogen depends both on the
temperature and the pressure within the atmosphere, the white dwarf
effective temperature and its surface gravity are strongly correlated
parameters, and acceptable fits are found for a 15000-25000\,K white dwarf,
corresponding to masses of 0.57-0.97\,M$_{\odot}$.
The best fit was achieved for a  
0.97M${\odot}$, 25,000K white dwarf (Figure 8). Even at this temperature,
the bluest wavelengths are not fit
well, implying some other component may be contributing.  

Although the {\it GALEX} and optical data
were not simultaneous, we tried to constrain any hot component by combining the
closest datasets.
Figure 9 shows the CTIO data from April 17 plotted with the {\it GALEX}
data from May 13 (fluxes are plotted in frequency to show the data better),
with the observed range of variability of the {\it GALEX} data. 
The UV data does not follow the upturn at the bluest
optical wavelengths. A hot 25,000K model lies far above the {\it GALEX}
points. This could mean that either there is a large reddening (not
very feasible given the small distance of GW Lib), or there was a large
change in the {\it GALEX} or optical fluxes between the two measurements
(also not very feasible), or the model parameters
are not correct. Using the best determination of mass from
van Spaandonk et al. (2010) i.e.  mass of 0.84 M${\odot}$ and the distance
of 104 pc with a small amount of reddening (E(B-V)=0.03) yields the
fit shown in Figure 9 for a temperature of 19,000K. In this model, the
optical sum of the white dwarf and disk falls below the observed values.
However, this could be possible if the disk contributes a greater portion
of the optical continuum flux. The fitting of UV spectra have shown that
the white dwarfs in accreting pulsators contribute 75-89\% of the UV light
and 42-75\% of the optical (Szkody et al. 2010) during quiescence. 
In the model shown in
Figure 9, the white dwarf contributes 80\% of the flux near 4500\AA.
Since GW Lib was not yet at quiescence, the disk contribution could
be larger than normal. 

Given all the uncertainties (10\% in optical
flux calibration, 20-30\% in distance, unknown changes in the month between the {\it GALEX} and CTIO
observation times, unknown reddening and unknown disk contribution and
flux distribution), it is not worth pursuing a more detailed fit at this
time. It is clear that there is a hotter component (likely the white dwarf
as the dominant contributor to the flux) than quiescence a year
after its outburst. In WZ Sge, the white dwarf was about 4000K hotter than
quiescence one year after its outburst (Godon et al. 2006). Given the
similar physical characteristics of the two systems, including both having
massive white dwarfs, it is interesting to determine if 
GW Lib remains hotter at the same
time post
outburst than WZ Sge. Since the cooling is related to the amount and
depth of the accreted mass, the larger outburst amplitude of GW Lib (9
mags) vs WZ Sge (7.5 mags) may indicate that more material was accreted
during the outburst of GW Lib. In addition, the much lower inclination
of GW Lib provides a better view of the disk radiation at outburst.
The determination of the actual cooling of the white dwarf in 
GW Lib compared to WZ Sge
will come from ultraviolet spectra. 

A soft X-ray component was seen with Swift (Byckling et al. 2009) 
within a few days of outburst that was consistent with an optically thick 
boundary layer. The hard X-ray flux was abnormally large at outburst
(1000 times its pre-outburst value) and was still an order of magnitude
larger than pre-outburst through 2009 March. Whatever caused the
increased X-ray luminosity could result in a slower cooling
of the white dwarf in GW Lib compared to WZ Sge.

\section{Conclusions}

Our UV and optical data on GW Lib during the three years following its
outburst has shown the following unique traits:

\begin{itemize}

\item The accretion disk shows large optical 
variability following the steep 
decline part of its post-outburst light curve.

\item A superhump modulation evident only in optical, not ultraviolet, 
follows this transition.

\item One year post-outburst, a 19 min quasi-period is visible in the optical 
for 2-4 months.

\item The ultraviolet begins to show variability on $\sim$4 hour timescales a year after
the outburst 
and the amplitude of this variability 
increases in the second and third years after outburst.

\item The optical begins to show a 4 hr variation 2 years after
outburst which is similar to the ultraviolet but not in phase and this variation
can disappear from one night to the next.

\item At one year past outburst, the white dwarf likely remained hotter than
at quiescence, although a range of temperatures and disk components is
possible. 

\item There is no evidence of the return of the pre-outburst pulsations
for the 3 years following outburst.

\end{itemize}

The very large outburst amplitude, the long-lasting X-ray emission, the hot
white dwarf, the
lack of post-outburst rebrightenings, the presence of quasi-periodicities and the
long period modulation all point to a large
accretion episode, resulting in a hot white dwarf
and a massive accretion disk. While the structure of the disk that
results in all the periodicities observed is not known, following the
detailed transitions in systems like this provide some clues to the types
of phenomena that the disk undergoes as it transitions from its dominant optically
thick emission state to its minor quiescent contribution to the system light.
Obtaining the pulsation periods and amplitudes when the white dwarf cools
enough to resume its modes will reveal the effect of the accretion event
on the interior layers of the white dwarf.

It is interesting to compare the behavior of GW Lib after its outburst
to that of two other similar systems that also had recent outbursts and
contain pulsating white dwarfs. SDSS0745+45 had an outburst in October 2006,
showed no visible pulsations one year post outburst (Szkody et al. 2010) but
pulsations at the same pre-outburst periods had returned by February 2010 
(Mukadam et al. 2010b). V455 And
had an outburst in September 2007 and its pulsation return is currently
being monitored. A comparison of the outburst amplitudes and return timescales
and characteristics of the pre-and post outburst pulsations of these three systems
should provide some clues to the depth of heating and interaction of the
accretion event with the driving zone of the pulsations.
\acknowledgments

This research was funded by NASA GALEX grants NNX08AU43G, NNX09AF87G
  and NSF grant AST0607840.

\clearpage

\begin{deluxetable}{cccll}
\tablewidth{0pt}
\tablecaption{Summary of ${\it GALEX}$ Data}
\tablehead{
\colhead{UT Date} & \colhead{mid UT} & \colhead{Exp (s)} & \colhead{FUV} & \colhead{NUV}
}

\startdata
2007 May 26 & 18:19:55 & 1088 & 14.439$\pm$0.004 & 14.874$\pm$0.004\\
  & 18:58:36 & 1120 & 14.446$\pm$0.004 & 14.840$\pm$0.004\\ 
  & 21:37:16 & 1136 & 14.420$\pm$0.004 & 14.818$\pm$0.004\\
2007 Jun 08 & 13:50:20 & 106 & 14.70$\pm$0.01 & 15.07$\pm$0.01\\
2007 Jun 18 & 03:40:31 & 221 & 14.79$\pm$0.01 & 15.16$\pm$0.01\\
  & 05:19:05 & 616 & 14.799$\pm$0.007 & 15.165$\pm$0.006\\
  & 06:57:41 & 1002 & 14.805$\pm$0.006 & 15.179$\pm$0.005\\
  & 08:36:16 & 1388 & 14.806$\pm$0.005 & 15.192$\pm$0.004\\
  & 10:14:53 & 1690 & 14.780$\pm$0.004 & 15.180$\pm$0.004\\
2008 May 13 & 17:52:59 & 1321 & 15.948$\pm$0.009 & 16.179$\pm$0.007\\
  & 19:31:39 & 1343 & 16.046$\pm$0.009 & 16.222$\pm$0.007\\
  & 21:12:16 & 1201 & 15.864$\pm$0.008 & 16.110$\pm$0.006\\
  & 22:51:21 & 1197 & 15.968$\pm$0.009 & 16.136$\pm$0.007\\
2008 Jun 03 & 11:50:28 & 1684 & 15.986$\pm$0.007 & 16.175$\pm$0.006\\
  & 13:29:03 & 1667 & 15.917$\pm$0.007 & 16.142$\pm$0.006\\
  & 15:07:39 & 1678 & 16.074$\pm$0.008 & 16.261$\pm$0.006\\
2008 Jun 13 & 10:05:33 & 1683& 16.071$\pm$0.008 & 16.276$\pm$0.006\\
  & 11:44:09 & 761 & 16.02$\pm$0.01 & 16.21$\pm$0.02\\
2009 May 16 & 10:41:23 & 1498 & 16.23$\pm$0.008 & 16.31$\pm$0.005\\
  & 12:19:59 & 1498 & 15.88$\pm$0.007 & 16.03$\pm$0.004\\
  & 13:58:36 & 1503 & 16.27$\pm$0.008 & 16.34$\pm$0.005\\
2010 Apr 25 & 03:35:27 & 1488 &  & 16.35$\pm$0.005\\
  & 05:14:03 & 1489 & & 16.13$\pm$0.004\\
  & 06:52:40 & 1460 & & 16.50$\pm$0.005\\
  & 08:35:13 & 1471 &  & 16.04$\pm$0.004\\
\enddata
\end{deluxetable}
\clearpage

\begin{deluxetable}{llllccl}
\tabletypesize{\footnotesize}
\tablecaption{Summary of Optical data}  
\tablehead{
\colhead{UT} & \colhead{Site} & \colhead{JD-2450000} & \colhead{UT range} & 
\colhead{Exp (s)} & \colhead{Filters} & \colhead{Comments}
}
\startdata
2007 Apr 16 & NMSU & 4206 & 08:50:03-11:02:15 & 60 & $U$ & \\
2007 Apr 17 & NMSU & 4207 & 06:59:51-10:23:26 & 4.1-219.5 & $U$ & \\
2007 Apr 18 & NMSU & 4208 & 07:01:08-10:58:14 & 3.4-46.7 & $UBVRI$ & DFT in $U$ \\
2007 Apr 19 & NMSU & 4209 & 07:02:39-09:50:54 & 3.7-93.6 & $UBVRI$ & DFT in $U$
\\
2007 Apr 20 & NMSU & 4210 & 06:46:42-10:49:32 & 3.7-47.9 & $UBVRI$ & DFT in $U$
\\
2007 Apr 22 & NMSU & 4212 & 09:05:40-10:23:21 & 3.5-50.7 & $UBVRI$ & DFT in $U$ \\
2007 Apr 26 & NMSU & 4216 & 07:24:36-07:32:44 & 2.0-33.4 & $UBVRI$ & \\
2007 Apr 27 & NMSU & 4217 & 06:52:53-07:01:26 & 2.5-44.0 & $UBVRI$ & \\
2007 May 04 & NMSU & 4224 & 06:09:03-06:18:07 & 3.6-58.9 & $UBVRI$ & \\
2007 May 07 & NMSU & 4227 & 05:37:41-09:40:43 & 1.4-219.8 & $UBVRI$ & DFT in $U$
\\
2007 May 10 & NMSU & 4230 & 05:50:58-09:54:46 & 3.0-138.7 & $UBVRI$ & DFT in $U$
\\
2007 May 11 & NMSU & 4231 & 08:42:35-09:19:50 & 3.1-85.0 & $UBVRI$ & DFT in $U$
\\
2007 May 13 & NMSU & 4233 & 08:17:08-09:33:15 & 4.3-162.7 & $UBVRI$ & DFT in $U$
\\
2007 May 14 & NMSU & 4234 & 07:52:53-09:25:18 & 2.2-107.4 & $UBVRI$ & DFT in $U$
\\
2007 May 29 & WIYN & 4249 & 05:43:44-07:57:11 & 15 & BG39 & \\
2007 Jun 01 & APO  & 4252 & 08:09:56-09:02:56 & 10 & BG40 & \\
2007 Jun 05 & NMSU & 4256 & 05:01:46-05:13:30 & 9.4-92.8 & $UBVRI$ & \\
2007 Jun 06 & NMSU & 4257 & 04:55:34-06:36:60 & 21.6-228.4 & $UBVRI$ & \\
2007 Jun 08 & NMSU & 4259 & 03:32:41-05:20:45 & 4.4-91.4 & $UBVRI$ & \\
2007 Jun 14 & NMSU & 4265 & 03:47:07-06:41:49 & 8.2-155.5 & $UBVRI$ & \\
2007 Jun 15 & NMSU & 4266 & 03:59:38-07:13:52 & 3.1-124.8 & $UBVRI$ & \\
2007 Jun 18 & NSMU & 4269 & 05:30:39-05:40:28 & 7.5-63.1 & $UBVRI$ & \\
2007 Jun 19 & NMSU & 4270 & 04:59:26-05:09:59 & 5.3-103.0 & $UBVRI$ & \\
2008 Mar 29 & APO  & 4554 & 07:56:43-11:25:53 & 10  & BG40 & \\
2008 Apr 17 & CTIO & 4573 & 07:08:19-08:09:19 & 2$\times$1800 &  & spectra \\
2008 Jun 13 & NMSU & 4630 & 03:51:41-07:03:03 & 6.7-136.6 & $UBVRI$ & DFT in $U$
\\
2008 Jun 20 & NMSU & 4637 & 04:01:36-06:16:10 & 3.0-112.3 & BG40,$UBVRI$ & DFT in BG40 \\
2009 May 14 & NMSU & 4965 & 23:21:59-01:16:45 & 60 & BG40 & \\
2009 May 15 & NOFS & 4966 & 05:15:10-09:14:30 & 20 & $V$ & \\
2009 May 16 & NMSU & 4967 & 01:11:23-03:24:12 & 60 & BG40 & \\
2009 May 16 & SRO  & 4967 & 04:34:30-09:50:26 & 45 & clear & \\
2009 May 16 & NOFS & 4967 & 06:04:55-09:45:49 & 45 & $B$ & \\
2009 May 17 & NOFS & 4968 & 04:51:22-09:19:45 & 45 & $B$ & \\
2009 May 17 & SOAR & 4968 & 05:04:24-09:58:28 & 8.9 & $V$ & \\
2009 May 18 & SAAO & 4969 & 21:47:40-01:39:30 & 40 & $B$ & \\
2009 May 19 & NMSU & 4970 & 05:19:52-08:28:29 & 60 & BG40 & \\
2010 Mar 04 & KPNO & 5259 & 10:07:09-12:36:21 & 15 & BG39 & \\
2010 Mar 05 & KPNO & 5260 & 10:03:19-11:36:37 & 15 & BG39 & \\
\enddata
\end{deluxetable}

\begin{deluxetable}{llcccc}
\tabletypesize{\footnotesize}
\tablewidth{0pt}
\tablecolumns{11}
\tablecaption{Summary of Observed Periods and Limits}
\tablehead{
\colhead{UT} & \colhead{Data} & \colhead{N} & \colhead{3$\sigma$ (mma)} &
 \colhead{Periods (min)} & \colhead{Amp (mma)}
}
\startdata
070416 & NMSU & 28 & 14.0 & & \\
070417 & NMSU & 145 & 9.2 & & \\
070418 & NMSU & 199 & 6.9 & & \\
070419 & NMSU & 70 & 11.2 & & \\
070420 & NMSU & 192 & 5.6 & & \\
070422 & NMSU & 52 & 3.6 & & \\
070510 & NSMU & 96 & 49.1 & 80.1$\pm$1.3 & 100$\pm$9 \\
070514 & NMSU & 41 & 73.9 & 83.3$\pm$1.3 & 137$\pm$3 \\
070526 & GALEX NUV & 120 & 17.1 & 32.3$\pm$0.4 & 21$\pm$3 \\
070526 & GALEX FUV & 120  & 26.9 & & \\
070529 & WIYN & 311 & 12.8 & 82.9$\pm$0.9 & 52.8$\pm$1.5 \\
070601 & APO & 278  & 14.3 & & \\
070618 & GALEX NUV & 294 & 36.2 & 154$\pm$2 & 78$\pm$7 \\
070618 & GALEX FUV & 294 & 44.4 & & \\
080329 & APO & 1255 & 5.7 & 204$\pm$4, 85.2$\pm$1.8,19.08$\pm$0.03 & 39.5$\pm$0.8,14.7$\pm$0.7,24.9$\pm$0.8 \\
080513 & GALEX NUV & 88 & 33.6 & 327$\pm$9 & 67$\pm$6 \\
080513 & GALEX FUV & 88 & 52.5 & 291$\pm$9 & 96$\pm$10\\
080603 & GALEX NUV & 87 & 39.3 & 275$\pm$22 & 66$\pm$7 \\
080603 & GALEX FUV & 87 & 49.5 & 219$\pm$8 & 106$\pm$22 \\
080613 & GALEX NUV & 42 & 38.7 & 215$\pm2$7 & 55$\pm$18 \\
080613 & GALEX FUV & 42 & 48.6 & & \\
080620 & NMSU & 89 & 27.1 & 124$\pm$6 & 53$\pm$6 \\ 
090514-15 & NMSU & 78 & 34.3 & & \\ 
090515 & NOFS & 719 & 10.9 & 248$\pm$7 & 60$\pm$2 \\
090516 & NMSU & 90 & 47.7 & 308$\pm$108 & 144$\pm$50 \\
090516 & SRO & 187 & 49.2 & 293$\pm$10 & 128$\pm$5 \\
090516 & NOFS & 169 & 41.2 & 206$\pm$11 & 109$\pm$7 \\
090516 & GALEX NUV & 78 & 81.7 & 210$\pm$9 & 150$\pm$8 \\
090516 & GALEX FUV & 78 & 105.7 & 218$\pm$11 & 191$\pm$9 \\
090517 & NOFS & 200 & 27.6 & 283$\pm$30 & 78$\pm$11 \\
090514-17 & multi  & 1438 & & 192$\pm$3 & 54$\pm$2 \\
090518-19 & SAAO  & 239 & 8.0 \\
090519 & NMSU & 61 & 20.5 &  &   \\
100304 & KPNO & 132 & 13.9 & 79.0$\pm$3.5, 40.2$\pm$0.9 & 21$\pm$3, 23$\pm$3 \\
100305 & KPNO & 90 & 16.2 & & \\
100425 & GALEX NUV & 30.5 & 103 & 229$\pm$4 & 205$\pm$8  \\
\enddata
\end{deluxetable}

\begin{figure}
\figurenum {1}
\plotone{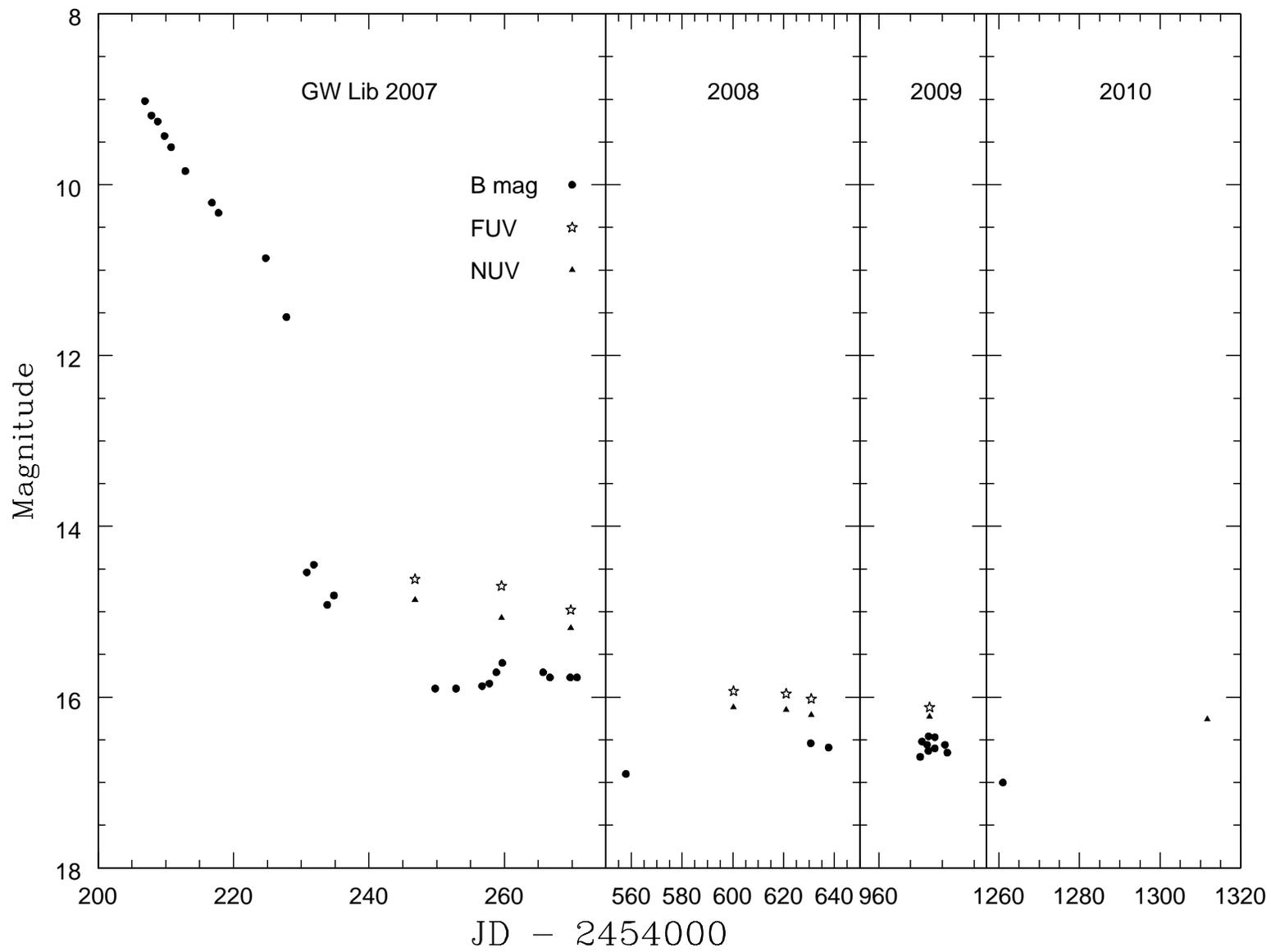}
\caption{Means of ultraviolet and optical light curves from 2007-2010.}
\end{figure}

\begin{figure}
\figurenum {2a}
\epsscale{0.9}
\plotone{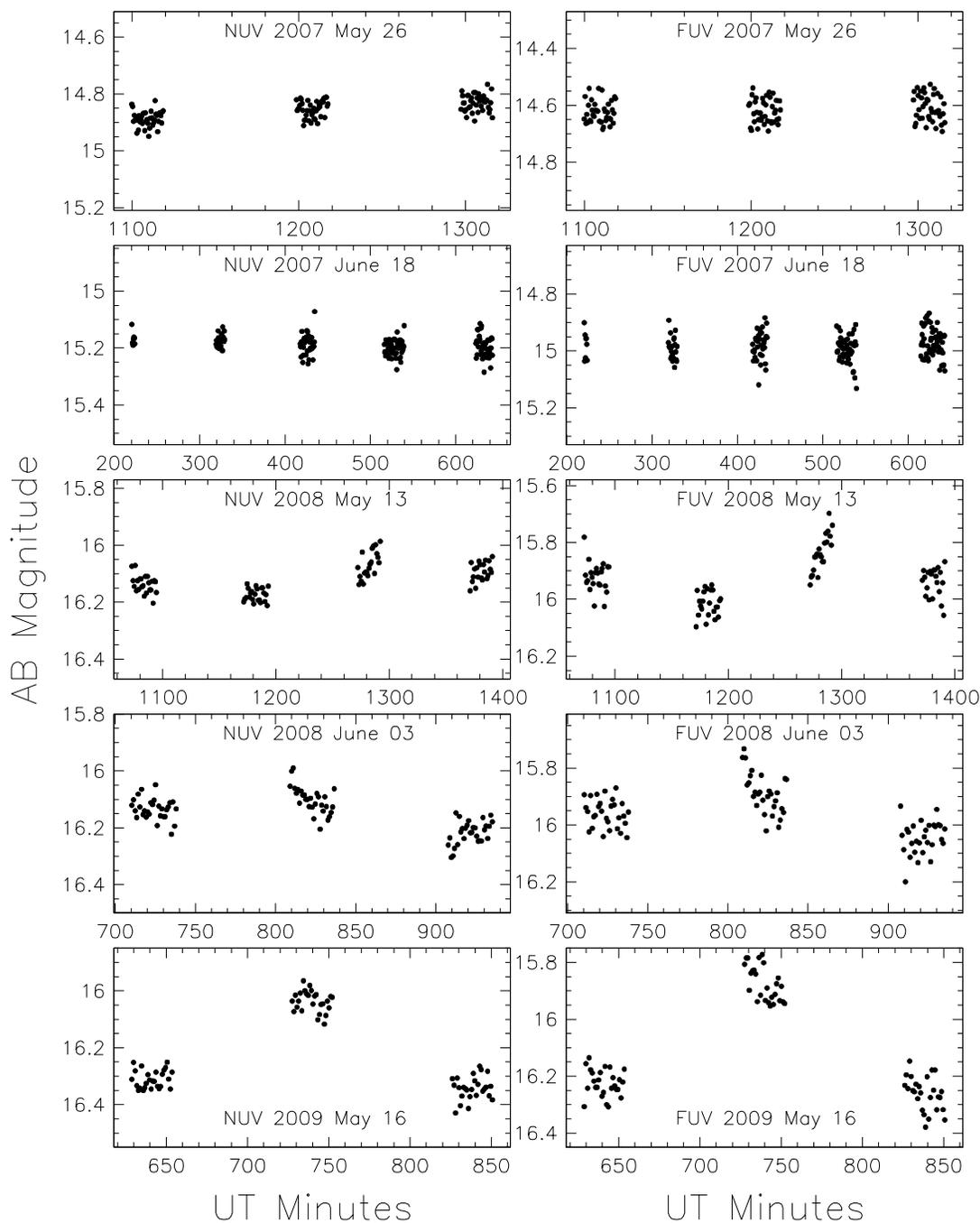}
\caption{{\it GALEX} NUV (left) and FUV (right) light curves. Each point
in 2007 data is a 29s integration, in later data is a 59s integration 
with error bars for each orbit listed in Table 1 . 
The magnitudes are on the AB system as explained in the text.}
\end{figure}

\begin{figure}
\figurenum {2b}
\plotone{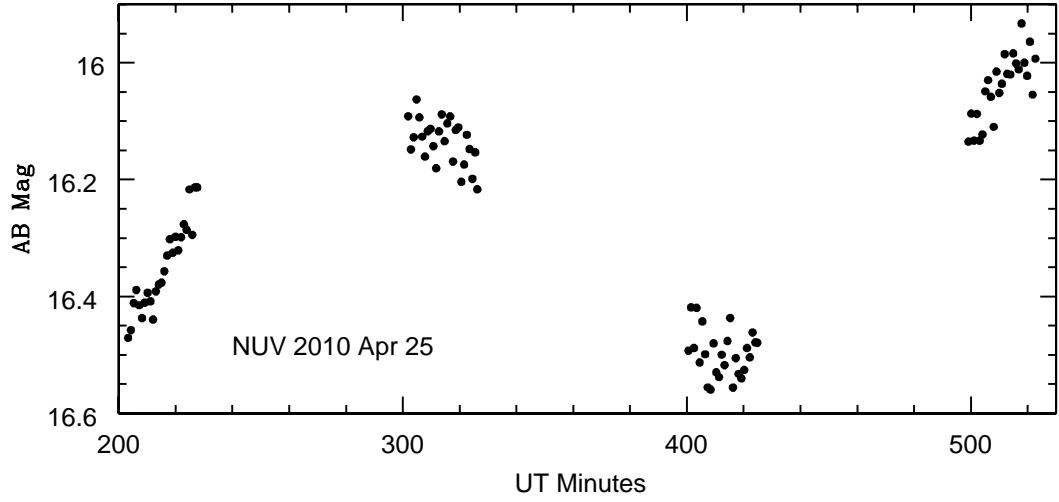}
\caption{{\it GALEX} NUV light curve in 2010. Details as in 2a.}
\end{figure}

\begin{figure}
\figurenum {3a}
\epsscale{0.9}
\plotone{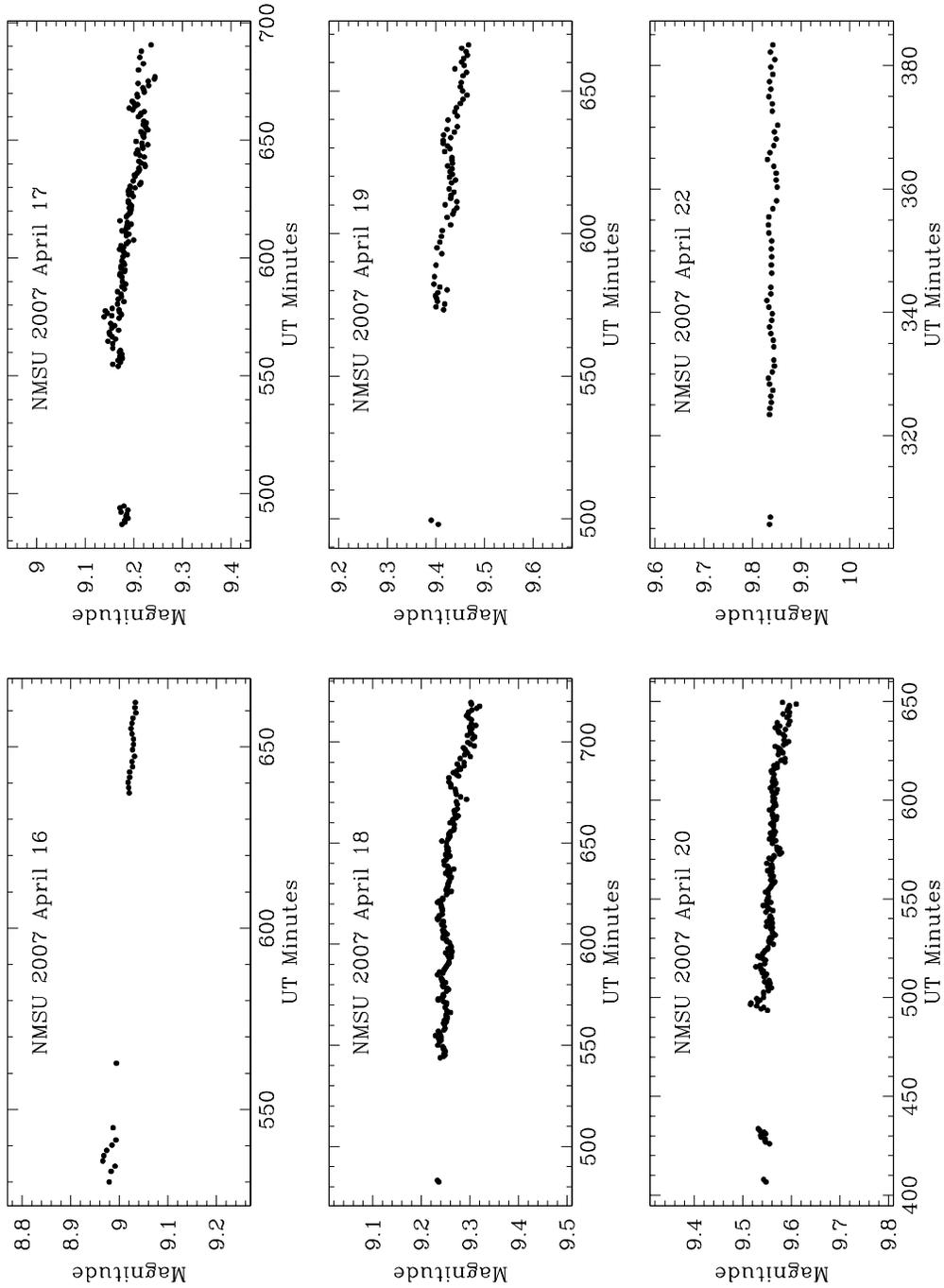}
\caption{Individual optical light curves obtained from 2007-2010. Error
bars are generally a few hundredths of a mag. Filters are as listed in Table
2, with filter same as for DFT if more than one on a night. Dates increase
across each row and then down the column.}
\end{figure}

\begin{figure}
\figurenum {3b}
\epsscale{0.9}
\plotone{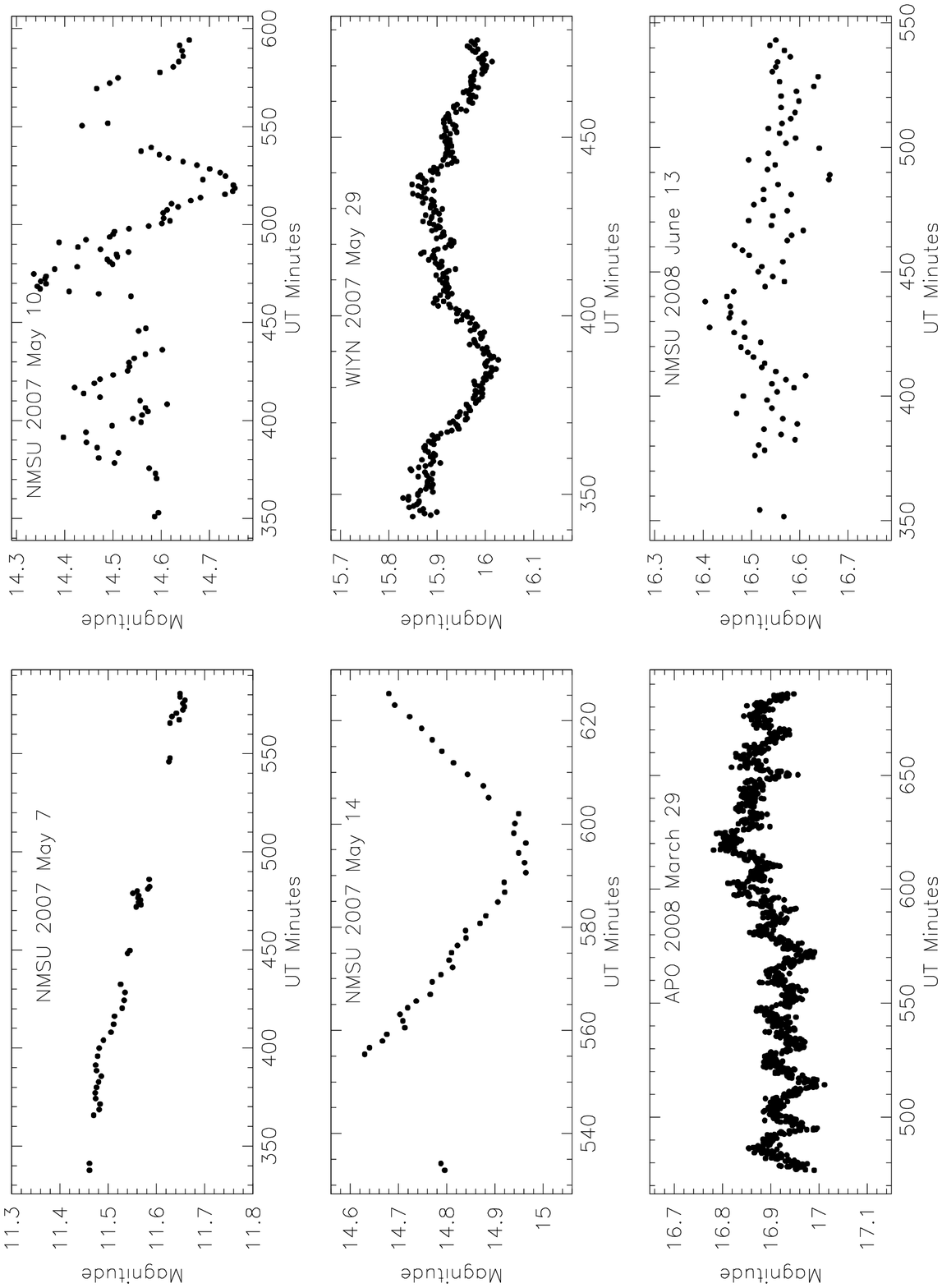}
\caption{Continued.}
\end{figure}

\begin{figure}
\figurenum {3c}
\epsscale{0.9}
\plotone{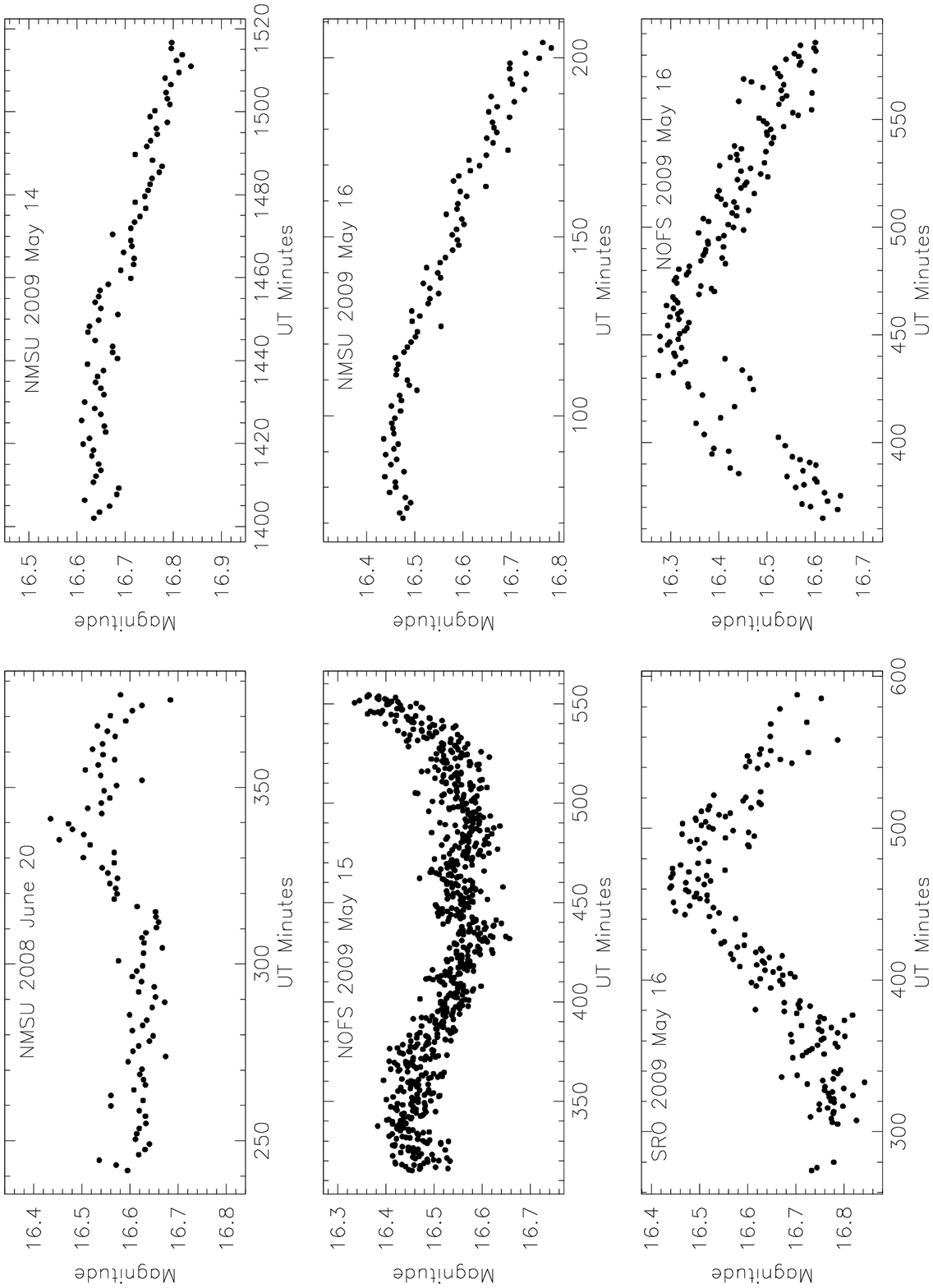}
\caption{Continued.}
\end{figure}

\begin{figure}
\figurenum {3d}
\epsscale{0.9}
\plotone{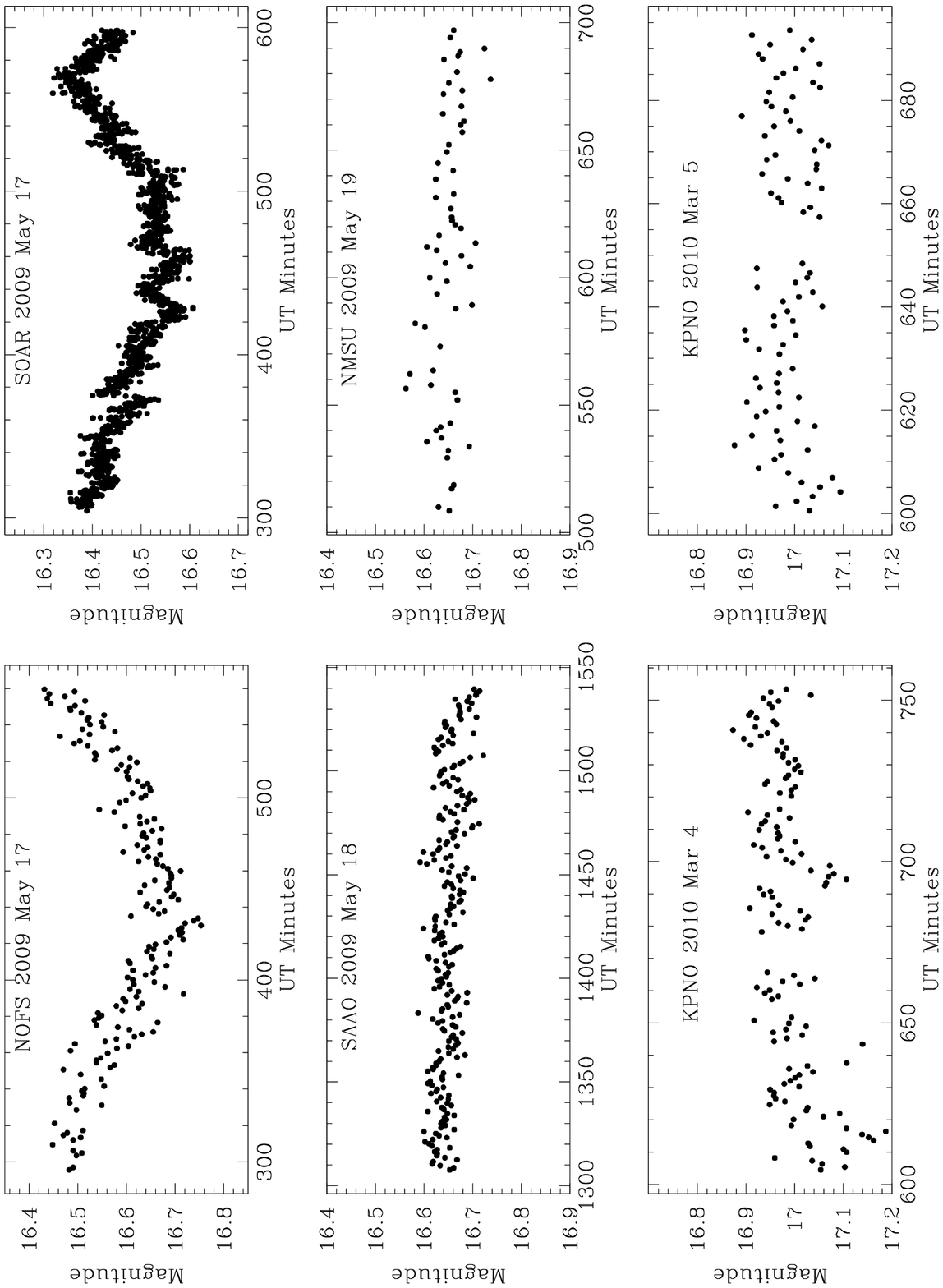}
\caption{Continued.}
\end{figure}

\begin{figure}
\figurenum {4}
\plotone{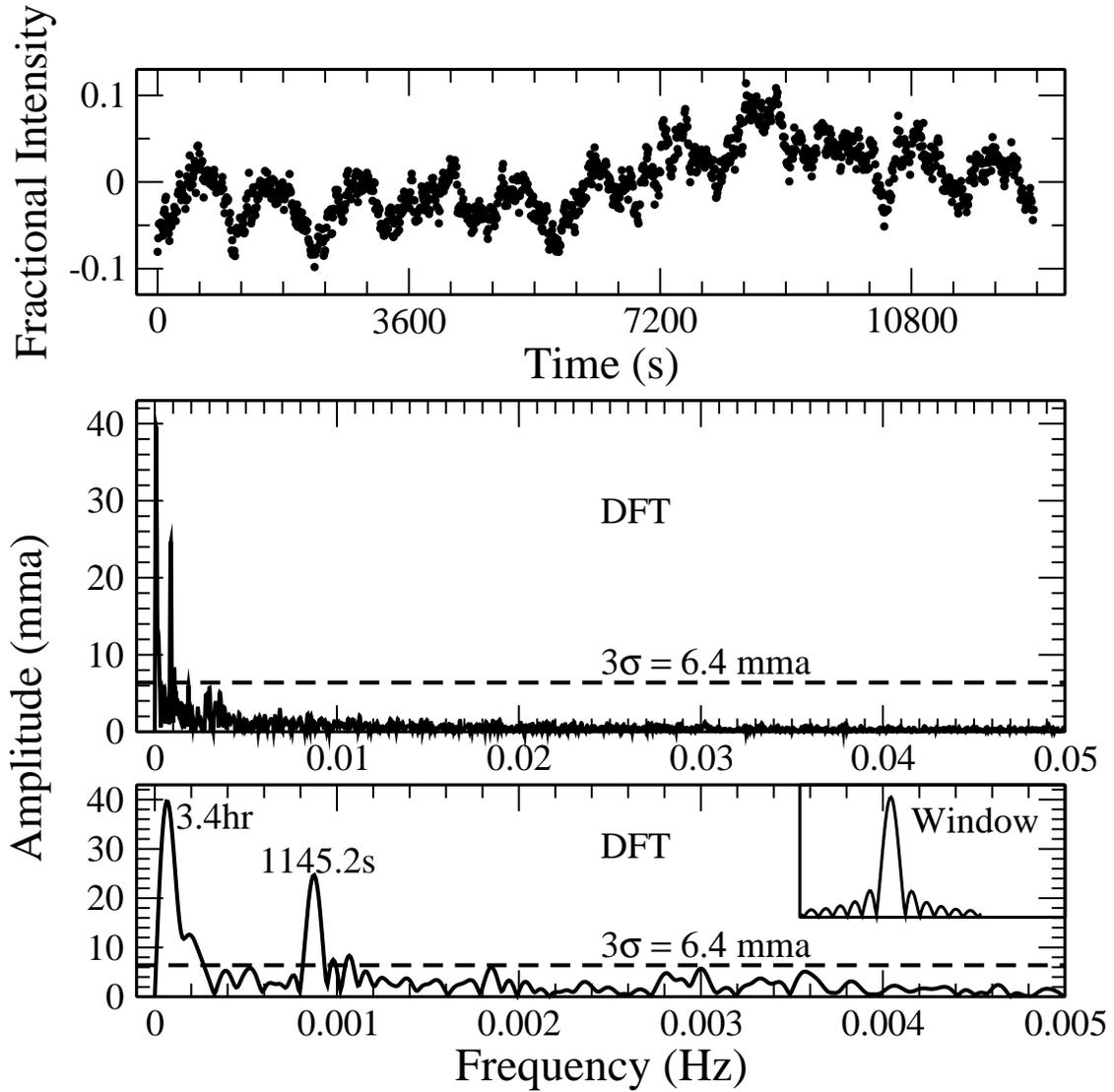}
\caption{Fractional intensity (top), DFT (middle) and enlargement of low 
frequencies (bottom) of APO optical data on 2008 March 29 showing the 19.08 min
 (1145s) period. 
Dashed line is the 3$\sigma$ noise limit.}
\end{figure}

\begin{figure}
\figurenum {5}
\plotone{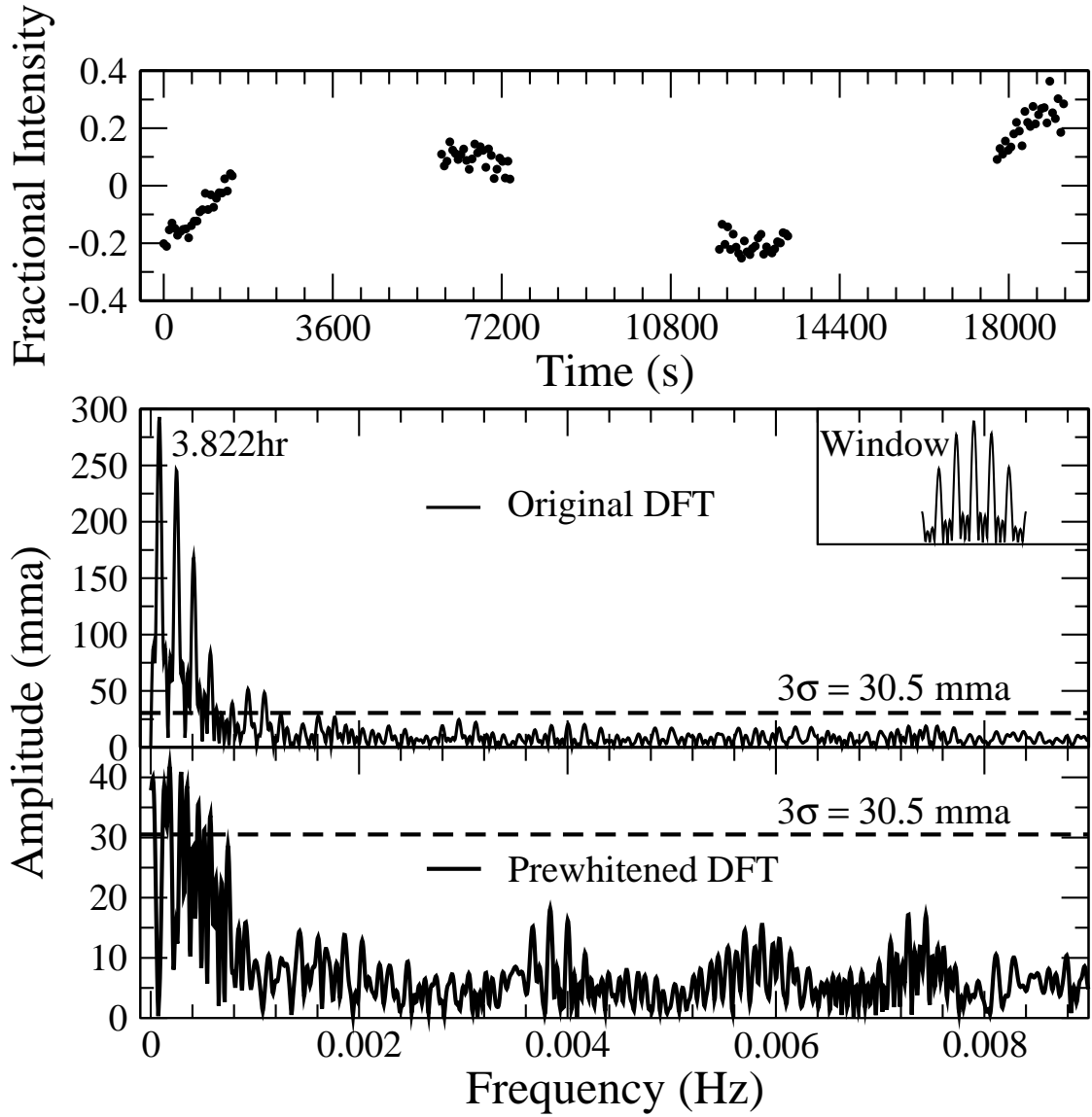}
\caption{Fractional intensity (top), DFT (middle) and DFT with 3.822 hr period
removed (bottom) of 2010 April 25 {\it GALEX} NUV data. Dashed lines show the 3$\sigma$ limits for the noise.}
\end{figure}

\begin{figure}
\figurenum {6}
\plotone{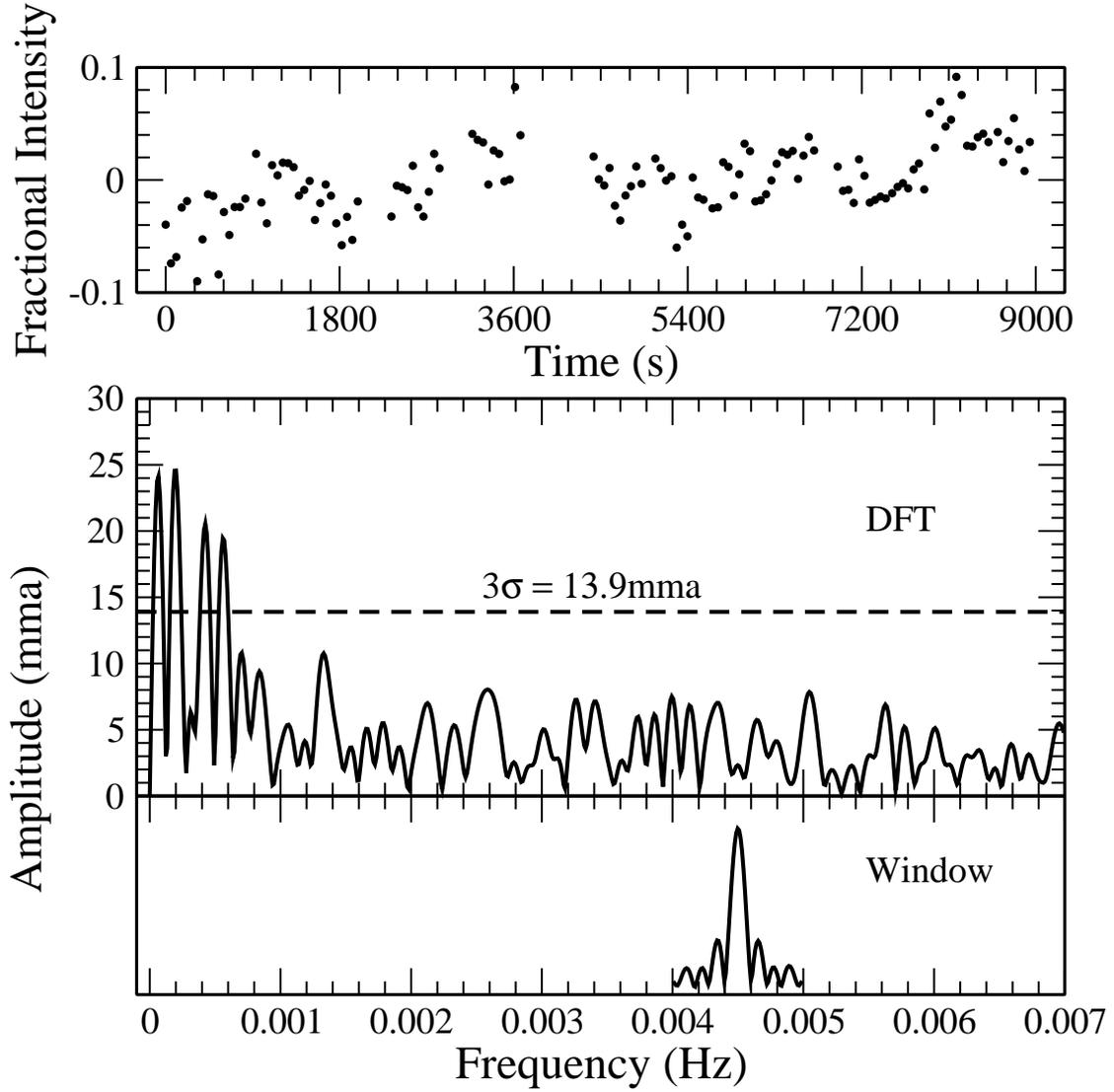}
\caption{Fractional intensity (top), DFT (middle) and window function (bottom) 
of 2010 March 4 optical data along with the 3$\sigma$ noise limit (dashed line).
The pre-outburst pulsations were all at frequencies larger than 0.001 Hz.}
\end{figure}

\begin{figure}
\figurenum {7}
\plotone{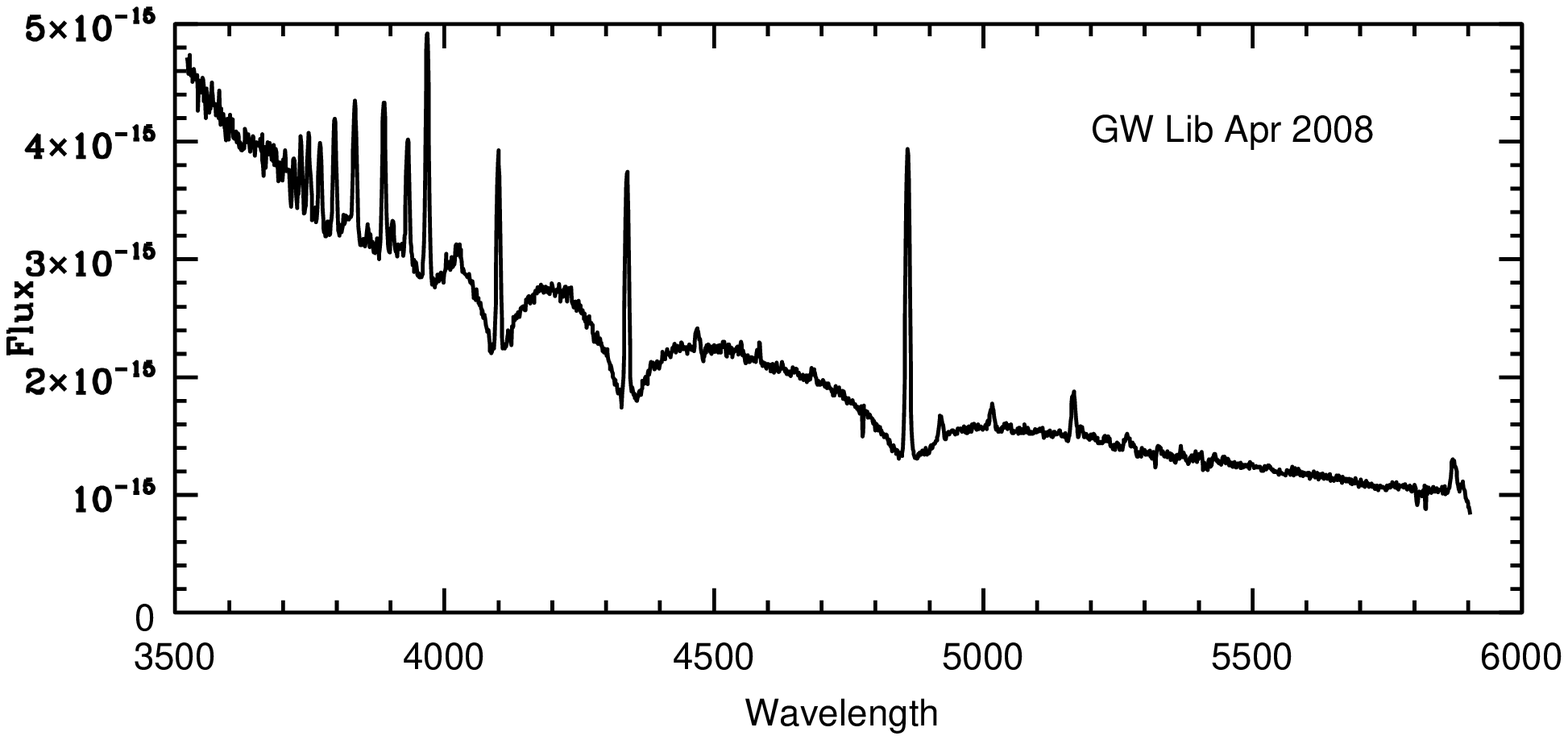}
\caption{CTIO spectrum of GW Lib. Note the broad absorption from the
white dwarf flanking the emission from the accretion disk.}
\end{figure}

\clearpage

\begin{figure}
\figurenum {8}
 \includegraphics [angle=-90,width=6in]{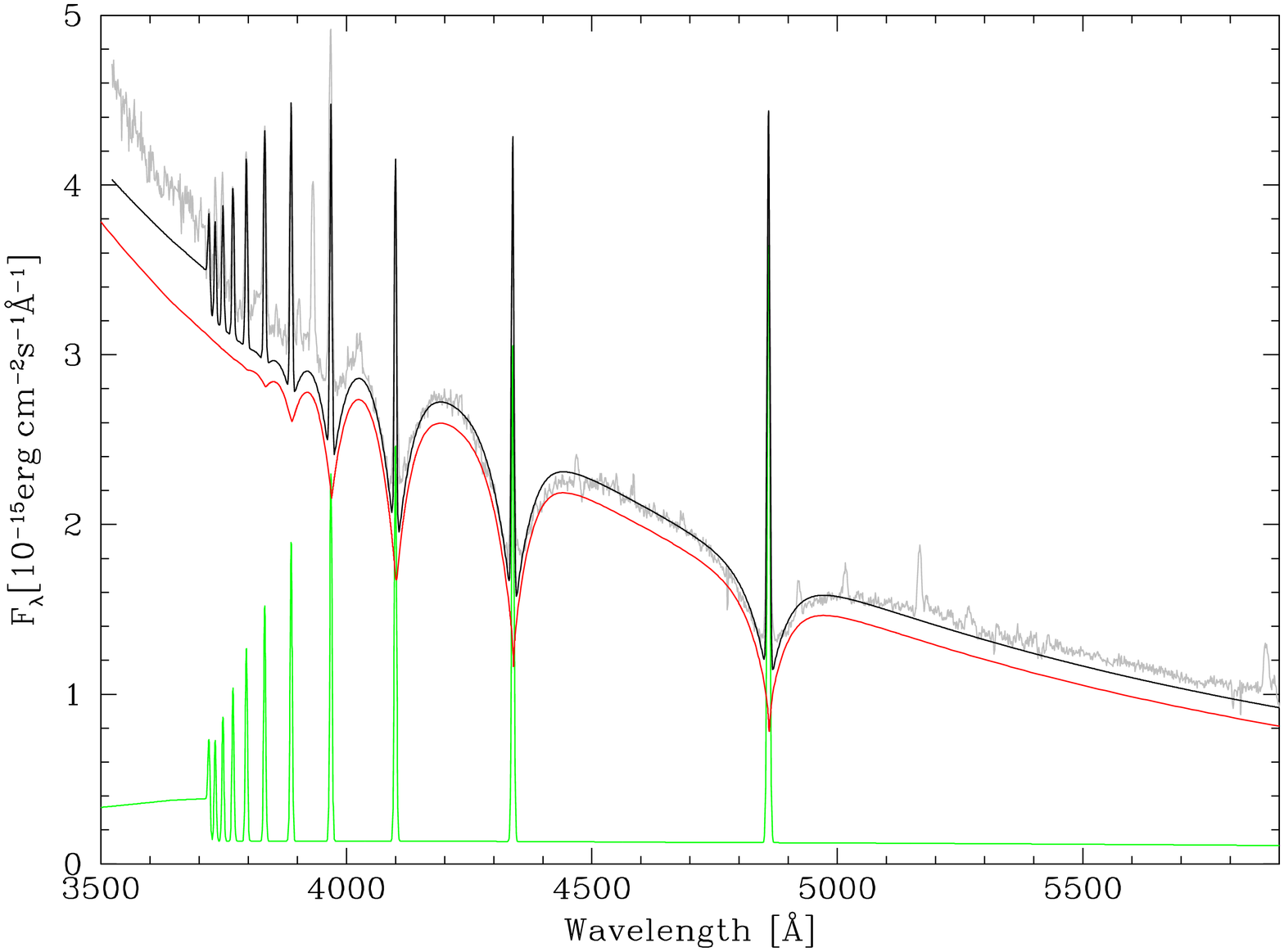}
\caption{CTIO spectrum (grey) fit with a 25,000K, log g=8.55 white dwarf at
104 pc (red) and an optically thin emission line component (green).} 
\end{figure}


\begin{figure}
\figurenum {9}
 \includegraphics [angle=-90,width=6in]{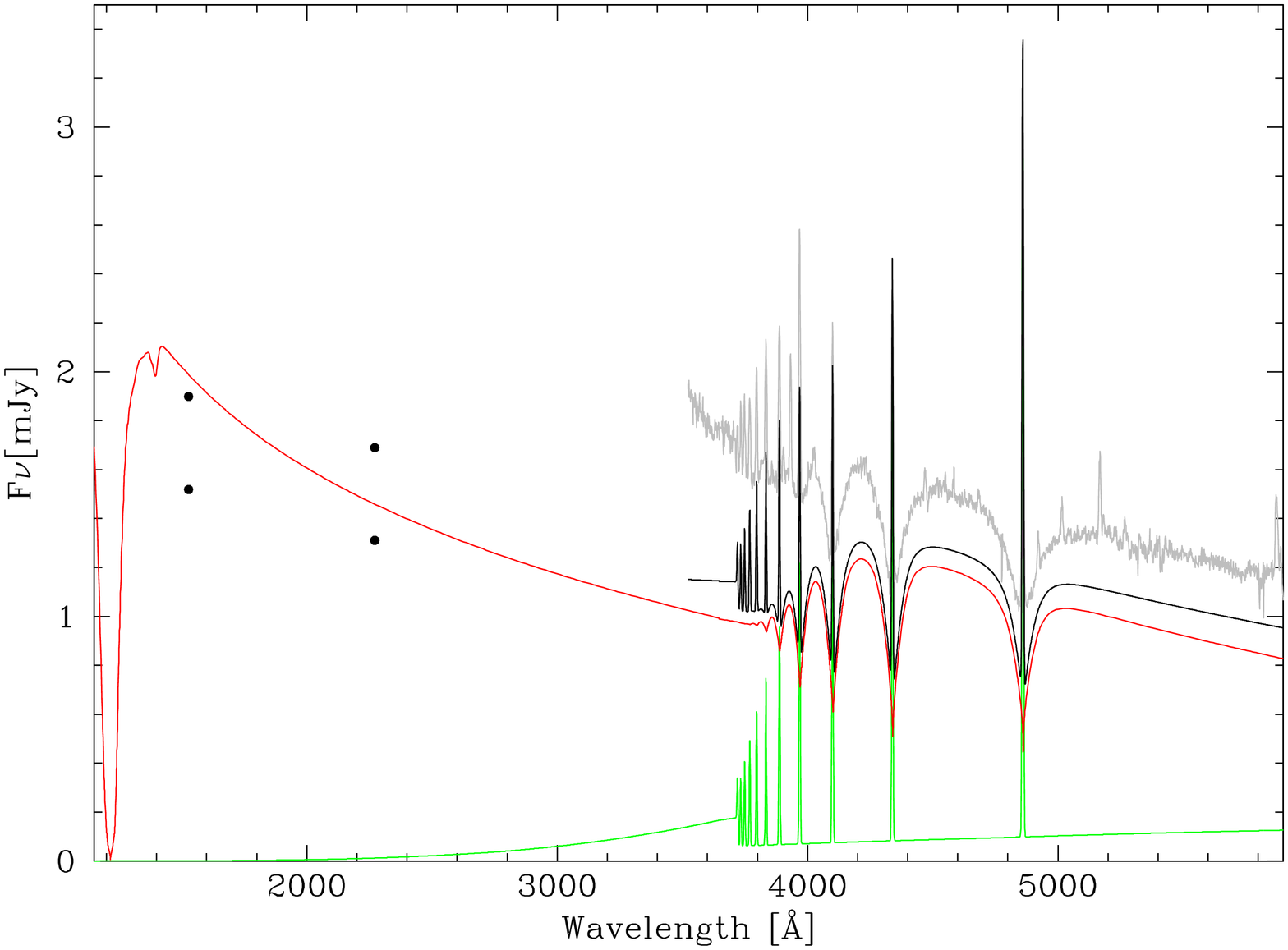}
\caption{{\it GALEX} data range from 2008 May 13 (solid points) 
combined with the CTIO spectrum from
 April 17 (grey) fit with a 19,000K, log g=8.36 white dwarf at 104 pc with a
reddening of E(B-V)=0.03 (red line) and an emission line component (green).}
\end{figure}


\begin{references}

\reference{} Araujo-Betancor, S. et al. 2005, \aap, 430, 629

\reference{} Andronov, I. L. 1999, \aj, 117, 574

\reference{} Byckling, K. et al. 2009, \mnras, 399, 1578

\reference{} Copperwheat, C. M. et al. 2009, \mnras, 393, 157

\reference{} Duerbeck, H. 1987, \ssr, 45, 1

\reference{} G\"ansicke, B. T. et al. 1997, \mnras, 289, 388

\reference{} G\"ansicke, B. T. et al. 1999, \aap, 347, 178

\reference{} Garnavich, P. \& Szkody, P. 1992, JAAVSO, 21, 81

\reference{} Godon, P., Sion, E. M., Cheng, F., Long, K. S., G\"ansicke, B. T. \& Szkody, P. 2006, \apj, 642, 1018

\reference{} Hamuy, M. et al. 1992, \pasp, 104, 533

\reference{} Hiroi,K. et al. 2009, \pasj, 61, 697

\reference{} Howell, S. B., Szkody, P. \& Cannizzo, J. 1995, \apj, 439, 337

\reference{} Hubeny, I. \& Lanz, T. 1995, \apj, 439, 875

\reference{} Kato, T., Maehara, H. \& Monard, B. 2008, \pasj, 60, 23

\reference{} Kepler, S. O. 1993, Baltic Astronomy, 2, 515

\reference{} Martin, D. C. et al. 2005, \apj, 619, L1

\reference{} Mukadam, A. S. et al. 2010a, \apj, 714, 1702

\reference{} Mukadam, A. S. et al. 2010b, Proc. 17th European White Dwarf Workshop,
AIP, in press

\reference{} Nogami, D. et al. 2009, ASPC, 404, 52

\reference{} Patterson, J. et al. 2002, \pasp, 114, 721

\reference{} Robinson, E. L. et al. 1995, \apj, 438, 908 

\reference{} Rodriguez-Gil, P. et al. 2005, \aap, 431, 269

\reference{} Schwieterman, E. W. et al. 2010, JSARA, 3, 6

\reference{} Steeghs, D. et al. 2007, \apj, 667, 442

\reference{} Szkody, P., Desai, V. \& Hoard, D. W. 2000, \aj, 119, 365

\reference{} Szkody, P.,  G\"ansicke, B. T., Howell, S. B. \& Sion, E. M. 
2002, \apj, 575, L79

\reference{} Szkody, P. et al. 1998, \apj, 497, 928

\reference{} Szkody, P., et al. 2010, \apj, 710, 64 

\reference{} Taylor, C., Thorstensen, J. R. \& Patterson, J. 1999, \pasp, 111, 184

\reference{} Templeton, M., Stubbings, R., Waagen, E. O., Schmeer, P.,
 Pearce, A., \& Nelson, P. 2007, CBET, 922, 1

\reference{} Thorstensen, J. R., Patterson, J., Kemp, J. \& Vennes, S. 2002,
 \pasp, 114, 1108

\reference{} Thorstensen, J. R. 2003, \aj, 126, 3017


\reference{} van Spaandonk, L., Steeghs, D., Marsh, T. R., Torres, M. A. P. 
2010a, \mnras, 401, 1857

\reference{} van Spaandonk, L., Steeghs, D., Marsh, T. R., \& Parsons, S. G. 2010b, \apj, 715, L109


\reference{} van Zyl, L., Warner, B., O'Donoghue, D., Sullivan, D., Pritchard, J
., \& Kemp, J. 2000, Baltic Ast., 9, 231

\reference{} van Zyl, L., et al. 2004, \mnras, 350, 307

\reference{} Warner, B. \& van Zyl, L. 1998, IAU Symp. 185, 321

\reference{} Whitehurst, R. 1988, \mnras, 232, 35

\reference{} Wood, M. A. 1995, Lecture Notes in Physics, 443, 41

\reference{} Woudt, P. \& Warner, B. 2002, Ap\&SS, 282, 433

\end{references}
\end{document}